# Generation of ultra-intense spatiotemporal optical vortex

*Renjing Chen*[1, 2, 3#], *Yilin Xu*[1, 2#], *Fengyu Sun*[1, 2], *Shunlin Huang*[3], *Xiong Shen*[3], *Wenpeng Wang*[1, 2], *Jun Liu*[1, 2, 3\*], *Ruxin Li*[1, 2, 3]

[1]*State Key Laboratory of Ultra-intense Laser Science and Technology, Shanghai Institute of Optics and Fine Mechanics, Chinese Academy of Sciences, Shanghai 201800, China*

[2]*University Center of Materials Science and Optoelectronics Engineering, University of Chinese Academy of Sciences, Beijing 100049, China*

[3]*Zhangjiang Laboratory, 100 Haike Road, Pudong, Shanghai 201210, China*

[#] These authors contributed equally.

\*Corresponding author: jliu@siom.ac.cn

**Abstract:** Spatiotemporal optical vortex (STOV) with transverse orbital angular momentum (TOAM) can induce some novel properties in high energy density physics. However, the current STOV pulse energy is limited to the mJ level, which greatly hinders the development of the research field of relativistic laser-matter interaction. Combined with the large-scale grating pair in high-peak-power laser facility, the method for generating of STOV with ultra-high intensity up to $10^{21}$ W/cm$^2$ is proposed. The numerical simulation proves that the wave packet with 60 fs duration and 83 J energy can be generated in the far field, maintaining an integral spatiotemporal vortex construction. Finally, STOVs with 1.1 mJ single pulse energy were obtained in a proof-of-principle experiment, and characterized by a home-made measuring device.

## 1. Introduction

Wave packet with vortex structure in spatiotemporal dimension is called spatiotemporal optical vortex (STOV)[1], which is being studied popularly in recent years. In view of its particular characteristic containing transverse orbital angular momentum (TOAM)[2], STOV has significant potential applications in high energy physics[3], high-speed information transmission[4] and quantum communication. Considering the importance of ultra-intense vortex laser in the field of relativistic laser-matter interaction[5, 6], the ultra-intense STOV will present novel phenomenon and great application value simultaneously. Although there have been extensive researches on STOV, such as its propagation characteristics[7-9], frequency transforms[10-13], manipulations[13-15] and characterizations[16, 17], the interactions between ultra-intense STOV and matters are still restricted in numerical simulation[18]. Recently, applications of STOV with relativistic intensity (usually > $10^{18}$ W/cm$^2$) were proposed and the relevant simulations had been conducted, including the generation of isolated attosecond electron sheet[19], γ photons and pairs[20] and γ-ray pulse[21], generation of intense spatiotemporal high-harmonic[22-25], electron acceleration and generation of attosecond hard x-rays[26], governing of interference photoelectron momentum distributions[27].

During the research of particle acceleration, STOV also plays an especial role in it. Traditional ultra-intense Gaussian lasers accelerate high-energy particle beams with large divergence angles due to their unique Gaussian distribution of pondermotive force[28, 29]. This is not conducive to many applications, such as fast ignition[30], the generation of high-brightness radiation sources[31, 32], and the excitation of high-energy nuclear physics[33]. The emergence of ultra-intense vortex lasers with spatial helical wavefront structures, such as Laguerre-Gaussian (LG) lasers, confine and modulate spatially high-energy particles, achieving the acceleration of collimated particle beams due to the spatial hollow pondermotive force distribution [34, 35]. Meanwhile, the longitudinal orbital angular momentum (OAM) of the LG laser can be transferred to the accelerated particle source and generated radiation source[36]. Nevertheless, such an LG laser can only achieve spatial modulation of the particle beam, but lacks temporal modulation, which limits the application and development of high-energy particle beams and radiation sources in ultrafast fields[37]. The emergence of the ultra-intense STOV laser is expected to solve this problem, because it has a hollow pondermotive force distribution in spatiotemporal dimensions. This is expected to extend the spatial modulation by the LG laser to the spatiotemporal modulation by the STOV laser[38]. At the same time, the transverse OAM carried by the STOV laser will be transferred to the particle beam, expanding a new degree of freedom on the basis of the LG laser.

However, the scheme generating the STOV with single pulse energy >1 mJ is infrequent, let alone the ultra-intense STOV with tens of Joule pulse energy. Now, there are two types of universal methods to generate STOV. One is based on the 4f system or analogous system[8, 39], sculpturing the spatiotemporal structure in Fourier plane by phase mask. This method is only suitable for the small-caliber (usually <1 mm) and narrow spectral width (usually <5 nm) for the limited propagation distance and limited caliber of cylindrical lens used to align the beam. At the same time, the damage threshold of grating is the dominant factor limit the energy of STOV below the mJ level. So, considering the restriction of spot size, spectral width and grating damage threshold, the traditional 4f system is not suitable for ultra-intense STOV generation. Another method is to directly generate STOV through micro-structured materials[40-44]. The intensity of STOV generated by this kind of method is also limited by the damage threshold of the materials. Meanwhile, the low efficiency and high cost hinder the practical application of this delicate component. In conclusion, neither 4f system or micro-structured materials owns the ability of producing ultra-intense STOV. Therefore, it is necessary to put forward a universal and robust method to generate STOV with several or even tens of Joules.

In this manuscript, a novel method by modifying grating compressor to generate ultra-intense STOV is proposed. The simulation results demonstrated that modified grating compressor system is able to generating STOV with tens of Joules energy, and $10^{21}$ W/cm$^2$ peak intensity can be achieved in the focus. With existing ultra-high peak power facility: Shanghai Ultra-intense Laser Facility (SULF), we can obtain STOV with above $10^{20}$ W/cm$^2$ peak intensity, just by modifying the structure of the grating based compressor in the chirped pulse amplification (CPA)[45]. Finally, a proof-of-principle experiment was conducted to verify this method, with a followed corresponding simulation results proving its practicability.

## 2. Method

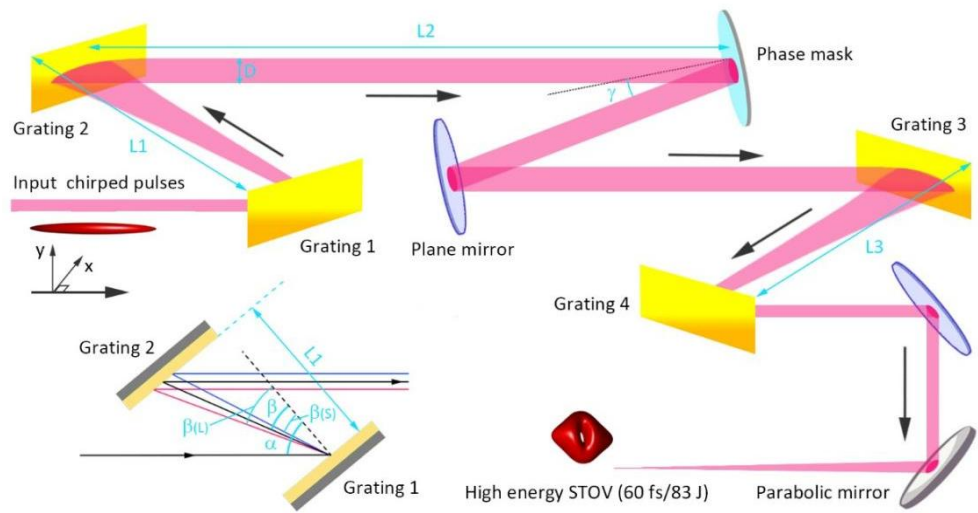

Figure. 1 Modified grating compressor for the generation of ultra-intense STOV with 60 fs pulse duration and 83 J pulse energy. L1 and L3 are the perpendicular distances of the first and the second grating pair, respectively. D is the diameter in X direction after the Grating 2. L2 is the distance between the Grating 2 and Phase mask. $\gamma$ is the incident angle of Phase mask which depends on the actual experimental condition.

Since that damage threshold of grating is the main factor limiting the peak intensity of STOV, it is nature to expand the spot-size of input beam to increase the pulse energy. Simultaneously, broader spectra means shorter duration and focal intensity. Adopting femtosecond light source with broadband spectrum is beneficial. Nevertheless, the key processes of generating STOV are the introduction of spatial dispersion. Broadening the spectra with adequate spatial dispersion will expand the diameter in X direction. In the case, large-sized cylindric lens with a diameter of more than 1 m is needed. In addition, wavefront distortion is unavoidable. Compared to the cylindric lens, grating pair has the same dispersion and alignment functions but without wavefront distortion. So far, the production of large-scale gratings is feasible.

Furthermore, grating pair can not only introduce spatial dispersion, but also negatively temporal dispersion. A symmetric grating pair can be set after the phase mask to compensate the equal spatial dispersion. Considering the negatively temporal dispersion caused by grating pairs, the input pulse should be positively pre-chirped.

Fortunately, for ultra-high peak power laser, the stretcher and compressor in CPA can induce pre-chirp to the input pulse and generating STOV easily, as Figure. 1 shown. The phase mask is placed between the two grating pairs to modulate the spatial and spectral phase of the laser. At last, STOV can be generated in the near or far field of the four-grating compressor STOV generation system (FGC-SGS), depending on the function in phase mask. Next, we will analyze the specific parameters of FGC-SGS, based on the Fourier diffraction theory and dispersive theory.

## 3. Numerical simulation

Based on above analysis, the spatial dispersion and beam alignment are achieved by using a grating pair. And the input pulse before the Grating 1 should be positively chirped.

The input pulse can be represented as

$$E_1(x,y,t) = \sqrt{I_0}\exp[-(x^2+y^2)^{10}/a^{20}]\exp[-(1+iC)t^2/2T_0^2] \qquad (1)$$

In formula (1), $I_0$ is the peak intensity, $a$ is the radius of spot, $C$ is the chirp amount, and $T_0$ is the temporal duration.

After the grating pair, the temporal and spatial dispersion introduced by it can be expressed with the following equation.

$$\phi(\omega,f_x) = \phi_0 + \phi'\Omega + \frac{\phi''}{2}\Omega^2 + \tau_x\frac{f_x}{k_0}\Omega \qquad (2)$$

Here, $\phi(\omega,f_x)$ is the relative phase delay of different temporal frequency $\omega$ and different transversely spatial frequencies $f_x$. $\phi'$ and $\phi''$ are the group delay and group delay dispersion (GDD)[46], respectively. $\Omega = \omega - \omega_0$ is the frequency difference between the each frequency component and central frequency $\omega_0$. The last term in the right is the delays of spatial frequency by time, which $\tau_x$ is the delay coefficient [47], and $k_0 = \frac{2\pi}{\lambda_0}$ is the wave vector in vacuum.

Therefore, the optical field after the grating pair is represented as:

$$E_2(x,y,t) = IFFT\{E_1(f_x,f_y,\omega)\cdot\exp[i\phi(\omega,f_x)]\} \qquad (3)$$

Note that $E_1(f_x,f_y,\omega) = FFT_2[E_1(x,y,\omega)]$ is the two-dimension Fourier transform in spatial domain for $E_1(x,y,\omega)$, and $E_1(x,y,\omega) = FFT[E_1(x,y,t)]$ is the Fourier transform in temporal domain. The optical field after the free propagation of distance L2 can be expressed as:

$$E_3(x,y,\omega) = IFFT_2\{E_2(f_x,f_y,\omega)\cdot H(f_x,f_y,\omega)\} \qquad (4)$$

where $H(f_x,f_y,\omega) = \exp[ikL - i\pi\lambda_0 L(f_x^2 + f_y^2)]$ is the free propagation function.

Then the optical field after the phase mask is
$$E_4(x,y,t) = E_3(x,y,t)\cdot P(x,y) \qquad (5)$$
in which $P(x,y)$ is the function of phase mask.

The following processes are the same with the previous steps (4) and (3) but with a reverse order. The optical field after the Grating 4 is $E_5(x,y,t)$. The function of a parabolic mirror can be approximately written as $F_{PM} = \exp[-\frac{ik_0(x^2+y^2)}{2f}]$, in which $f$ is the focal length. Finally, the STOV was formed at the focus of a parabolic mirror.

## 4. Results

### 4.1 Ultra-intense STOV generation

For a tenth-order super-Gaussian input pulse with radius $a$=115 mm, 800 nm central wavelength, 30 fs duration, (Note that if not specifically emphasize, pulse duration is the full width at half maximum: FWHM.) and 1480 line/mm gold plated gratings, employing a spiral phase mask, the STOV with topological charge $l$=1 is generated in the focal plane with $f$=0.5 m, as shown in Figure. 2. The intensity distribution presents a ringlike shape, with a phase singularity embedded in the Y-T plane, which is accorded with the description of STOV. A doughnut form iso-surface is shown in (c1), demonstrating that the direction of its intrinsic orbital angular momentum is along the X-axis, but not T-axis. More importantly, taking the damage threshold of gold grating as 0.2 J/cm$^2$, the maximal output energy can be up to 83.1 J in theory. Considering the spatial radius of STOV in focusing plane is about 3 μm, and the pulse duration is about 60 fs, it is easy to obtain the theoretically maximal intensity of $5\times10^{21}$ W/cm$^2$.

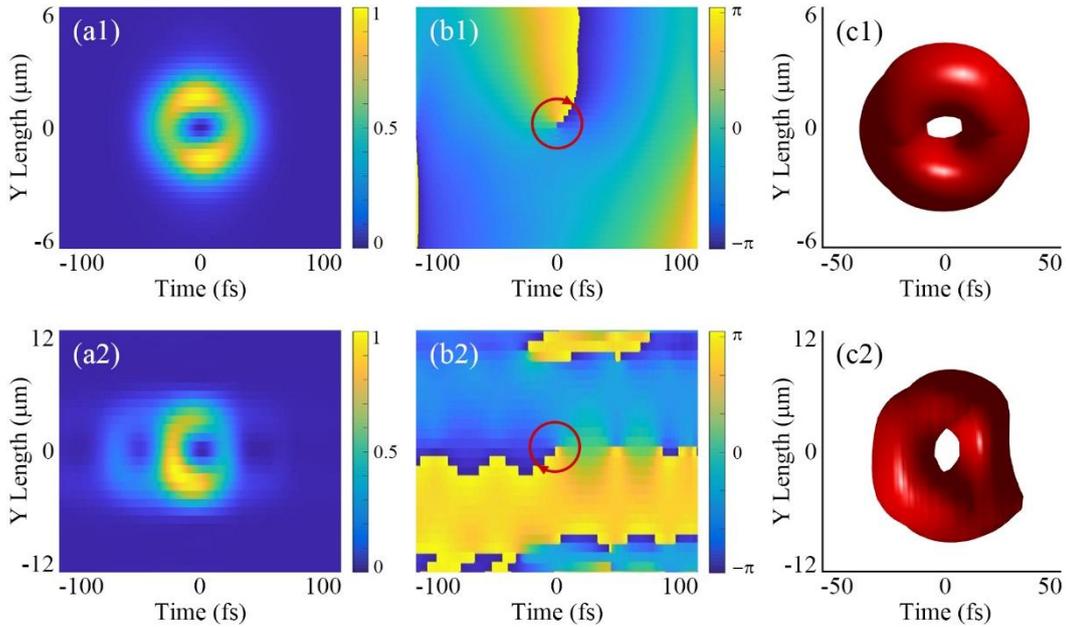

Figure. 2 Simulation results of the generation of ideal (a1)-(c1) and practical (a2)-(c2) ultra-intense STOVs in focal plane. (a1) and (a2) are the intensity distributions of STOV in Y-T plane at X=0 position. (b1) and (b2) are the corresponding phase distributions. (c1) and (c2) are the iso-surfaces at 1/3 maximum of intensity.

Although the STOV generated with four-grating compressor is convenient for most high power laser system, the specific optical path and component size should be taken into full consideration. To maintain enough spatial dispersion, the perpendicular distance of the grating pair should be at least 6 m. The result of it is the Grating 2 and Grating 3 have to be wider to adapt the dispersed spot. It is easy to calculate the diameter of output pulse from Grating 2 with formula (6).

$$D = 2a + [tan\beta_{(L)} - tan\beta_{(S)}] \cdot \cos(\alpha) \cdot L1 \qquad (6)$$

In formula, $\beta_{(L)}$ and $\beta_{(S)}$ are the diffraction angles of the longest and shortest wave components respectively, as shown in the embedded picture in Figure. 1. And the real spot width on grating surface is $D/\cos(\alpha)$. Assuming the incident angle $\alpha=49°$, grating period d=1/1480 mm, the corresponding grating width should be at least 1.32 m. Such a large-scale grating is under producing[48] and will be applied in the future ultra-high peak power laser facility.

On the other hand, the grating pairs introduced not only spatial dispersion but also temporal GDD, which is related to the grating pair perpendicular distance[46]. Therefore, it is necessary to calculate the duration of positively pre-chirped pulse before grating pair. If the generated STOV is a Fourier transform limit (FTL) pulse, the input pulse duration can be calculated with the following formula.

$$T_{input} = T_{FTL} \cdot [1 + \left(\frac{GDD \times 1.665^2}{T_{FTL}^2}\right)^2]^{0.5} \qquad (7)$$

For L1=L3=6 m, the $T_{input}$=3.00 ns.

Taking a real 1 PW facility: SULF in Shanghai as an example, the maximal grating size is 360 mm (H) × 565 mm (W), which limits the input beam size and perpendicular distance L1. If the proper beam diameter is set to 100 mm, the maximal L1 should be 2.6 m to make full use of the area of Grating 2. Substituting the practical laser parameters[49] in the simulation model, we can obtain the intensity and phase distribution of generated ultra-intense STOV as shown in Figure. 2(a2) and (b2). Besides the main spatiotemporal ring structure at T=0, there are several interconnected rings with low energy distributed on both sides, because of the temporal sidelobes in input pulse. And the iso-surface at focus is displayed in (c2), which is slightly different from (c1). Although the nonuniformity of intensity distribution in spatial domain and non-ideal spectrum distorted the STOV shape, it still maintained an intact vortex state. Besides, according to the above results, we can calculate the corresponding peak intensity of STOV, which is approximately 0.14×10$^{21}$ W/cm$^2$. And in temporal domain, the practical input pulse should be stretched to 1.29 ns to match the perpendicular distance L1 and L3.

### 4.2 Effect of phase mask function

As we all know, the function of phase mask directly decides the pulse structure in the focus. In the ultra-intense STOV generation process, the spot before the phase mask is expended more in X direction. Therefore, it is necessary to adjust the distribution ratio of the vortex phase in the X and Y directions to construct a perfect STOV shape. The phase mask function can be written as

$$P(x,y) = \exp[-il \cdot arctan(\sigma \cdot \frac{y}{x})] \qquad (8)$$

in which $\sigma$ is the distribution ratio coefficient. Similar to the previous simulation, the

STOV intensity and phase distributions in the focus with $\sigma$ from 0.2 to 5 are illustrated in Figure. 3. As the increase of the $\sigma$, the phase distribution range in X direction narrowed gradually, as shown from (a1) to (a5), resulting in a regular evolution from transverse double-lobes structure to circinate structure and finally to longitudinal double-lobes structure, as shown from (b1) to (b5). And the corresponding phase distributions are displayed from (c1) to (c5). Obviously, when $\sigma=2$, we can obtain the most appropriate spatiotemporal spiral structure. This is because for a large aperture spot, the spatially dispersed pulse should be uniformly modulated by spiral phase gradient. That means the function in phase mask should opportunely match the spot size on the mask. For a practical input pulse, the function is affected by the spot size, dispersive ability of grating and the distance L2, which can be strictly determined by simulation.

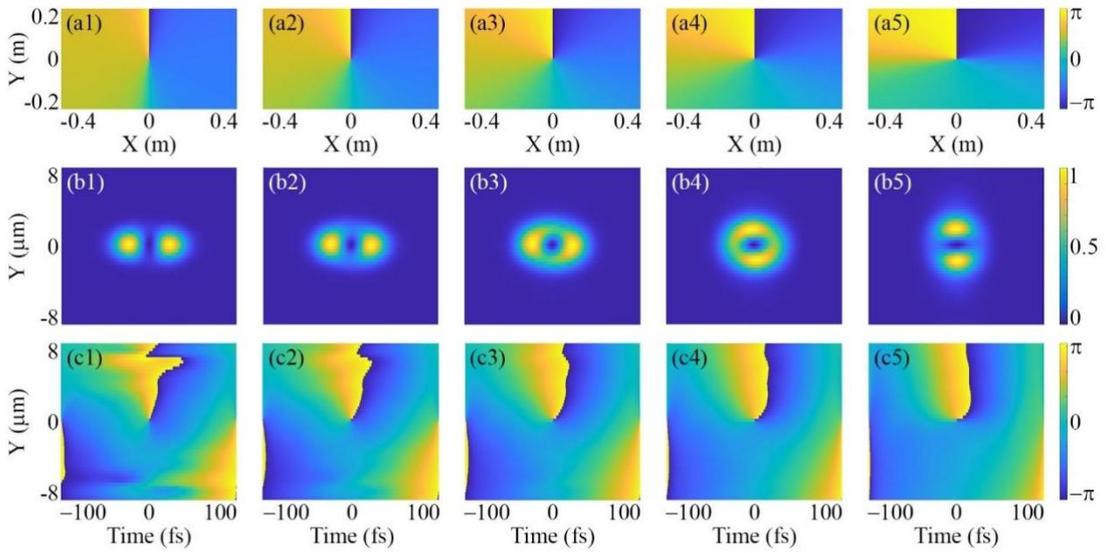

Figure. 3 (a1)-(a5)The different functions of phase mask, with $\sigma=0.2, 0.5, 1, 2, 5$ in sequence. (b1)-(b5) and (c1)-(c5) are the corresponding intensity and phase distributions in the focus, respectively.

**4.3 High-order ultra-intense STOV**

In the above text, we talked about the generation of ultra-intense STOV with topological charge $l=1$. In this part, high-order ultra-intense STOV with $l=2, 3, 4$ were produced in simulation, as shown in Figure. 4. All the STOVs were generated with $\sigma=2$. In the first line, it is easy to observe that the number of the hole in vortexes increased as the topological charge increasing. Simultaneously, the number of the phase vortexes also added in sequence, as shown in the second line. Due to the impact of phase singularities, the hole of the iso-surface enlarged as the increase of topological charge, but the whole wave packet still maintained a perfect donut shape, as shown in the third line.

Ultra-intense STOV with topological charge $l>1$ provide a new variable in the research of strong field physics, which can directly affect the TOAM of generated radiation source and accelerated particle source.

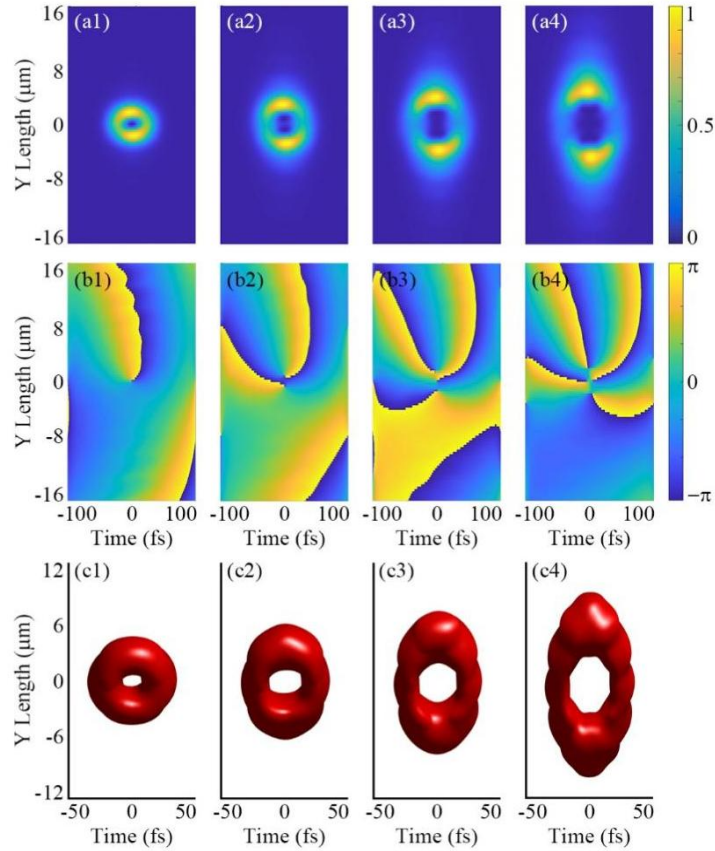

Figure. 4 The simulation results of STOV with topological charge $l$=1, 2, 3, 4, in turn. (a1)-(a4) and (b1)-(b4) are the intensity and phase distributions in the focus, respectively. (c1)-(c4) are the corresponding iso-surfaces at 1/3 maximum of intensity (The multi-hole structure is not appear in (c2)-(c4) due to the large iso-surface value).

## 5. Experiment

To verify the practicality of this ultra-intense STOV generation scheme, a miniaturization device was constructed in laboratory. The experimental schematic diagram is shown in Figure. 5. A 1 kHz repetition Ti: Sapphire CPA laser was used to produce the input pulse with 800-nm central wavelength, 120-ps pulse duration, 2.4 mJ energy and 6-mm spot diameter. The perpendicular distance L1 between the grating pair is 0.26 m. The distance from the second grating to the phase mask L2 equals to 0.5 m. Considering the measurement conditions, STOV pulses were generated in near field by a 1-inch phase mask having 0-π phase distribution[8]. The phase function of phase mask is as embedded picture in Figure. 6(a1) shown. The angle between the 0-pi dividing line and X axis is about 15 degrees. With two plane mirror, the pulses were reflected and backtracked to Grating 2 in different horizontal plane. Finally, the measurement of the generated STOV pulses was conducted with a home-made spectral interferometry with fiber array for single-shot spatiotemporal characterization

(SIFAST)[50]. The measuring results are illustrated in Figure. 6.

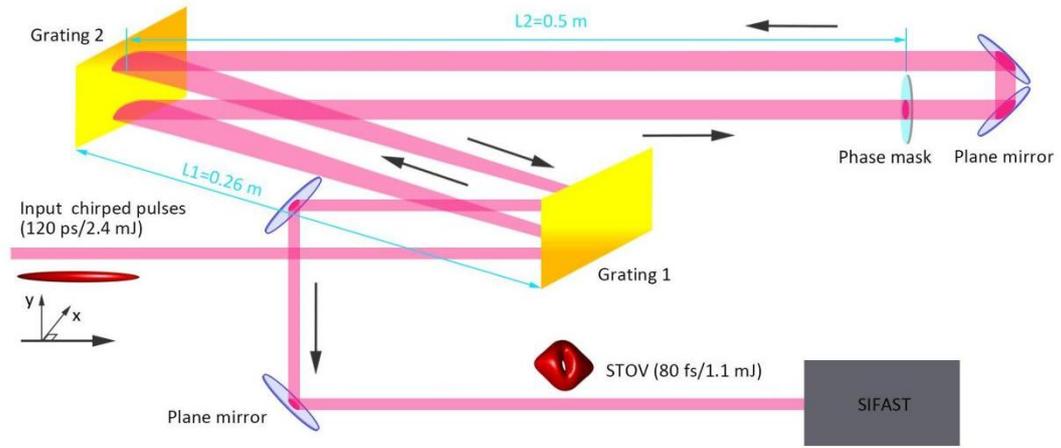

Figure. 5 Schematic diagram of the verification experiment.

The measuring resolution of SIFAST is 0.4 fs in temporal domain, and 100 um in spatial domain. And the measured temporal FWHM of the STOV is about 80 fs, which is closed to the simulation result. In Figure (a1), compared to the ideal iso-surface (a2), the experimental result shows nonuniformity in intensity distribution, because of the uneven spectral distribution and phase distortion. But the structure of singularity is clear in Y-T plane, accorded with the phase distribution in (b1). For phase distribution, benefited from the high resolution of SIFAST, the spiral phase was characterized distinctly, which is identical with the simulation one.

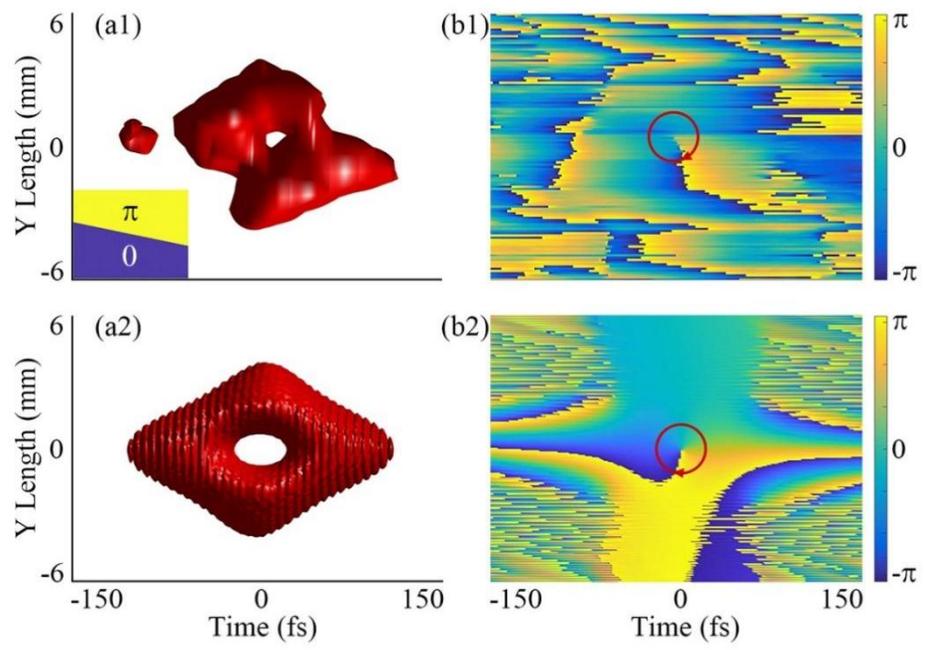

Figure. 6 (a1) and (b1) are the measuring results of iso-surface at 1/10 maximum of intensity and phase distribution in Y-T plane, respectively. The function of phase mask is embedded in the left bottom of (a1). (a2) and (b2) are the corresponding simulation results.

The 3D spectral data of the pulses collected by the SIFAST at intervals of 1.1 mm are shown in Figure. 7. The 13×12-point's spectra were plotted in Figure (a), as each group of spectra between the two dotted lines is the spectra of all the points in this row (Y1-Y13 rows). And the spectra of row Y7 were plotted detailedly in Figure. 7(b) (Y7 row, X1-X12 columns). Due to the 0-π function of phase mask, the boundary of phase step separate the spectra in both Y and X direction, as Figure (a) and (b) shown. Especially in row Y7, column X7, the spectrum was divided into two lobes equally, without spectral component of central wavelength. It is precisely this two part of spectra with a π phase difference that reconstitute a wave packet without dispersion after the second grating pair, but a phase singularity was retained at the center. Therefore, from the directly measured spectral data, we can also infer the characterization of the STOV is accurate and reliable.

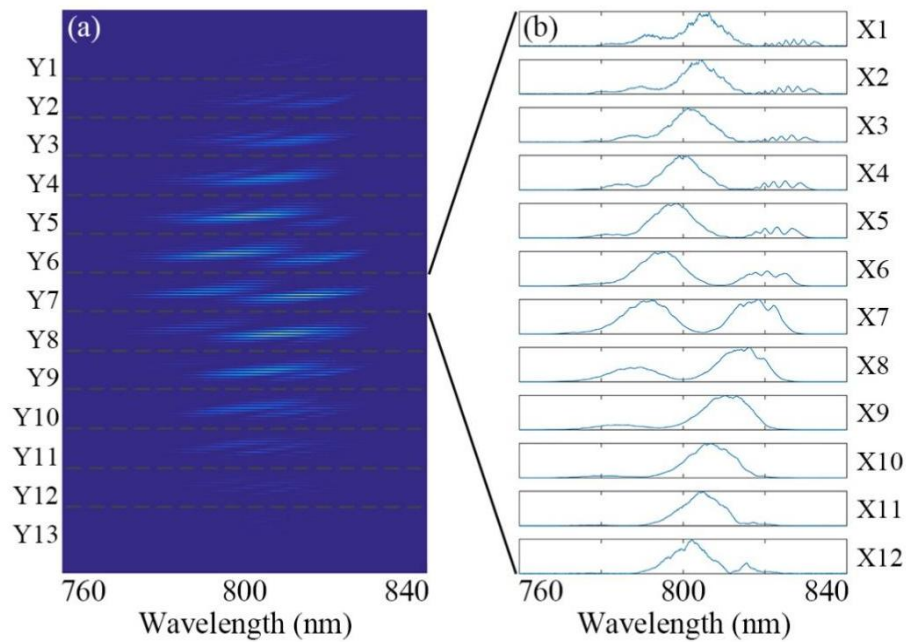

Figure. 7 (a) is the three-dimensional spectral interferometry map measured by spatiotemporal spectrometer in SIFAST. (b) are the spectra of X1 to X12 column, at Y7 line.

The single pulse energy of generated STOV was 1.1 mJ, with about 46% conversion efficiency. The main loss in generation system comes from the diffraction efficiency of the grating. So the calculated diffraction efficiency of one grating is 82%, which is the same as the laboratory measurement result. Actually, the energy is constricted by the laser's maximal output energy. In theory, it is possible to generate STOV with 50 mJ energy using the device in Figure. 5.

## 6. Conclusion

In this article, we first analyzed the application background of the ultra-intense STOV and its significance in strong field physics. Subsequently, we also introduced the potential methods for generating the ultra-intense STOV. Later, based on the ordinary 4f system, authors proposed the FGC-SGS method to generating ultra-intense STOV, which avoids the limitations of the grating damage threshold and the large-aperture cylindrical lens. It is the first time that an effective method was provided for the generation of STOV with tens of joule energy. In the third section, based on the Fourier diffraction formula and dispersive theory, authors conducted a simulation about the physical process of this method and obtained the ultra-intense STOV at the focus. The results showed that for a super-Gaussian pulse with 800-nm central wavelength and 230-mm diameter, theoretically a STOV with 60-fs duration, 83-Joule energy and $5 \times 10^{21}$ W/cm$^2$ peak intensity can be obtained at the focal point. Furthermore, considering the parameters of a practical ultra-high peak power laser facility: SULF, we obtained the information of an actual ultra-intense STOV that it could produce. Due to the limitations of the actual spot quality and pulse energy, the final peak intensity could be achieved was approximately $0.14 \times 10^{21}$ W/cm$^2$. The influence of the phase mask function and the generation of high-order ultra-intense STOV have also been thoroughly studied by simulations.

In the fifth section, a verification experiment was conducted with a small-size FGC-SGS device. About 1.1 mJ energy STOV was generated in the near field employing a 0-π phase mask with 46% conversion efficiency. The simulation results of the iso-surface and phase distribution were in good agreement with the experimental results, which once again proves the reliability and practicability of this method. On the other hand, no matter the ultra-intense STOV with tens of joule energy can be used in relativistic optics, STOV with mJ level energy can significantly promote the field of nonlinear effect of STOV, high-order harmonic STOV generation, and interaction between STOV and materials.

Overall, the FGC-SGS method does not require any additional optical components except for the phase mask. It can generate ultra-intense STOV simply by adjusting the existing four-grating compressor in high-energy laser facility, which is both cost-saving and efficient. The simple generation of the ultra-intense STOV will bring new research opportunities to laser-driven plasma physics.

**Data availability**
The data supporting the plots and other findings in this study are available from the corresponding author upon reasonable request.

**Code availability**
The simulation codes supporting the findings of this study are available from the corresponding author upon reasonable request.

**Acknowledgements**
This work was supported by the Shanghai Municipal Natural Science Foundation (No. 20ZR1464500), National Natural Science Foundation of China (NSFC) (Nos. 61905257 and U1930115), Shanghai Municipal Science and Technology Major Project (No. 2017SHZDZX02) and the Ministry of Science and Higher Education of the Russian Federation (Project No. FFUF-2024-0038).

**Author contributions**
J. L. and R. Ch. conceived the idea. R. Ch. performed the simulations. R. Ch., F. S. and Y. X. performed the experiments. J. L., R. Ch., F. S., Y. X. and X. Sh. analyzed the data. R. Ch. and F. S. prepared the manuscript and discussed it with all authors. J. L., W. W., S. H. and R. L. supervised the project.

**Competing interests**
The authors declare no competing financial interests.

Communications, **2025**.